\documentclass{ws-mpla}
\usepackage[super]{cite}
\usepackage{graphicx}

\usepackage{comment}
\usepackage{epsf}
\usepackage{color}
\usepackage{footnote}

\begin{document}



\title{Search for neutrinoless double beta decay}

\author{\footnotesize Igor Ostrovskiy}
\address{Physics Department, Stanford University, Stanford\\
California 94305,
USA\\
ostrov@stanford.edu}

\author{\footnotesize Kevin O'Sullivan}
\address{Physics Division, Lawrence Berkeley National Laboratory, Berkeley\\
California 94720,
USA\\
kevinosullivan@lbl.gov}

\maketitle

\pub{Received (Day Month Year)}{Revised (Day Month Year)}

\begin{abstract}
We review current experimental efforts to search for neutrinoless double beta decay (0$\nu\beta\beta$). A description of the selected leading experiments is given and the strongest recent results are compared in terms of achieved background indexes and limits on effective Majorana mass. A combined limit is also shown. The second part of the review covers next generation experiments, highlighting the challenges and new technologies that may be necessary to achieve a justifiable discovery potential. A potential synergy with direct dark matter searches, which could be an especially prudent strategy in case the axial vector coupling constant is quenched in 0$\nu\beta\beta$ decay, is emphasized.

\end{abstract}

\ccode{PACS Nos.: 14.60.Pq, 23.40.-s}

\section{Introduction}	

Neutrinos are assumed to be massless in the Standard Model as no right-handed neutrinos have ever been observed. Nevertheless, the neutrino flavor oscillation experiments unambiguously demonstrated that neutrino has a non-zero mass. The success of neutrino oscillation experiments notwithstanding, they are only sensitive to the mass squared differences, so we still do not know the absolute mass scale of neutrino.
It is also not known how the three mass eigenstates are aligned with respect to each other - the question of \textit{normal} versus \textit{inverse} hierarchy. Essentially, this question is whether the lightest neutrino mass state is dominated by electron neutrino flavor (normal), or by muon and tau neutrino flavors (inverted). 
Finally, it is not clear how exactly the mass term should be incorporated into the theory. Strongly linked to
the last question is the possibility that neutrino is its own anti-particle, which would make it the only known Majorona fermion.

The above unknowns could be explored if there exists a particular type of radioactive transition - the neutrinoless double beta decay (0\(\nu\beta\beta\)). Such a process would violate the lepton number conservation and B-L conservation, and is forbidden in the Standard Model. Among several possible mechanisms the simplest and most often considered one is the 0\(\nu\beta\beta\) decay mediated by a light Majorana neutrino with the emission of two electrons only. Another mode of this process that also often receives experimental attention is the 0\(\nu\beta\beta\) decay with an additional emission of one or two hypothetical bosons, called Majorons. Other potential mechanisms include exchanges by heavy Majorana neutrino and by sterile neutrino. Regardless of the mechanism, the existence of 0\(\nu\beta\beta\) decay would prove that neutrino Majorana mass is non-vanishing~\cite{Schechter:1982}. However, the relative contribution of the Majorana term to the total neutrino mass may still be negligible, as discussed in~\cite{Lindner}. In the simplest case of the exchange by a light Majorona neutrino, the half-life of 0\(\nu\beta\beta\) decay depends on the effective Majorona mass as follows:
\begin{equation}
T^{-1}_{1/2} = G^{(0)}_{0\nu}g^4_A|M_{0\nu}|^2|\frac{\langle m_{\beta\beta}\rangle}{m_e}|^2,
\label{half-life}
\end{equation}
where $G^{(0)}_{0\nu}$ is the phase-space factor, as defined in~\cite{iachello_psf}, that depends on the total energy available in the decay (Q-value) and details of the kinematics, $M_{0\nu}$ is the nuclear matrix element. $m_e$ is electron mass, $\langle m_{\beta\beta}\rangle$ is the effective Majorana mass of the electron neutrino, defined as a sum of neutrino mass eigenstates, $m_i$, weighted by corresponding elements of the neutrino mixing matrix, $U_{ei}$:
\begin{equation}
\langle m_{\beta\beta}\rangle = \sum_{i}U^2_{ei}m_i,
\label{mass}
\end{equation}
where the CP-violating phases were absorbed in the parametrization of the mixing matrix~\cite{MPLA_theory_review:2015}. However, the dependence of the half-life on the neutrino mass would be different for other mechanisms. To measure the neutrino mass scale from observation of 0$\nu\beta\beta$ would require knowledge of the dominant mechanism. This would likely involve observation in multiple isotopes or tracking of the full kinematics. The process of light-neutrino exchange would still have to contribute though, and assuming that non-observation implies a limit on the mass the electron-neutrino can have were it a Majorana particle is still considered safe.
Another problem of measuring the neutrino mass using this approach is the large theoretical uncertainties on the nuclear matrix elements (NME). Several approximations are used to calculate the NME resulting in differences by a factor of $\sim$2-4, depending on the isotope (see Fig.5 in~\cite{BilenkyGuinti:2015}). 

Recently another source of theoretical uncertainty is gaining attention and may result in serious repercussions for the next generation experimental searches. As is known from comparisons of $\beta$ and 2$\nu\beta\beta$ decay experiments with theory, the axial vector coupling constant $g_A$ needs to be renormalized in nuclear models~\cite{iachello_ga}. In particular, $g_A$ values may be decreased ("quenched") from its vacuum value $g_A\sim1.27$ down to $\sim$0.8~\cite{simkovich}. If the axial vector coupling is similarly quenched in 0$\nu\beta\beta$ decay, then a given experimental constrain on half-life will translate into 6-34 times weaker constraint on effective Majorana mass than currently assumed. 

\begin{figure}[ht]
\centerline{\includegraphics[scale=0.4]{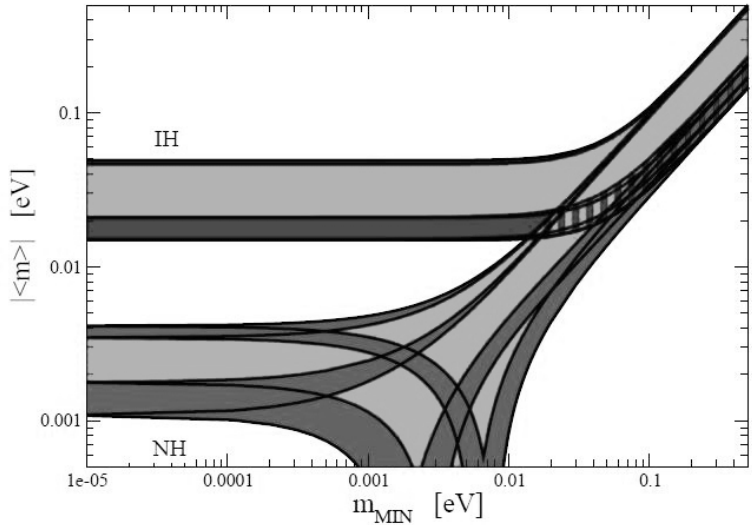}}
\vspace*{8pt}
\caption{The effective Majorana mass as a function of the lightest mass eigenstate ($m_1$ for normal hierarchy, NH, and $m_3$ for inverted hierarchy, IH). The bands include 2$\sigma$ uncertainty on the oscillation parameters. The light grey region corresponds to at least one of the Majorana CP violating phases having a non zero value, while dark grey region corresponds to CP conserving values. Adopted from~\cite{pdg:2014}.\protect\label{fig:eff_vs_min}}
\end{figure}

Dependence of the effective Majorana mass on neutrino oscillation parameters~\eqref{mass} defines the specific structure of the allowed parameter space, often expressed as a function of the lightest neutrino mass eigenstate, assuming the simplest mechanism of decay. Fig.~\ref{fig:eff_vs_min} shows that if the mass hierarchy is inverted, the effective Majorana mass can not be less than $\sim$15-20 meV, which provides a natural aim for the next generation experiments. Next generation neutrino experiments aim to answer the question of normal versus inverted hierarchy, and if the mass hierarchy is inverted next generation 0$\nu\beta\beta$ experiments would have the opportunity to make a discovery or rule out the process completely.

This paper gives brief overview of experimental searches for 0$\nu\beta\beta$ decay. It first describes the current state of the art experiments that are already probing the upper right corner of the allowed parameter space, called quasi-degenerate, due to relative similarity of all mass eigenstate values. The second part of the review covers proposed next generation experiments that aim to explore the inverted hierarchy (IH) region, highlighting the challenges, potential synergy with direct dark matter searches, and new technologies that may be necessary to achieve a justifiable discovery potential.

\section{Current state of the field}

A curious consequence of an apparent anti-correlation between the phase-space factors and the nuclear matrix elements for different double beta decay isotopes, noted in~\cite{robertson}, is that all isotopes are roughly equally attractive from the theoretical point of view. In that situation practical considerations dominate the isotope choice and the competition between existing and contemplated experimental efforts is being carried out along the three following axes:
\begin{enumerate}
\item exposure (isotope of interest mass times live-time)
\item radiopurity 
\item background rejection
\end{enumerate}
These three factors are difficult to optimize together and different technologies make different compromises.
For example, experiments using gaseous xenon may achieve background rejection significantly superior to experiments using liquid xenon (by being able to resolve detailed topological signature of $\beta\beta$ events and having better energy resolution), but the same exposure is more difficult to achieve  due to lower density of gas.
A commonly used combination of the three parameters in a single semi-quantitative estimate of experimental sensitivity\footnote{Usually defined as median expected 90\% C.L. half-life limit assuming no signal.} suggests that an experiment's limit on 0$\nu\beta\beta$ decay half-life scales as T\(_{1/2}\sim\sqrt{\frac{M\cdot t}{B\cdot \Delta E}}\), where $M$ is mass, $t$ is measurement time, $B$ is background rate in counts per unit mass per unit time, and $\Delta E$ is energy resolution. We want to stress, however, that singling out energy resolution as a way to maximize background rejection only makes sense if an experiment does not have any other source of background rather than unavoidable 2$\nu\beta\beta$ decay of the same isotope. 
In reality, the denominator is simply supposed to represent the total background in the region-of-interest (ROI) around the Q-value, so the energy resolution doesn't need to appear in this equation at all. 
A second reason we consider this equation antiquated is that some experiments don't have a flat background near their Q-value and instead may have a peak close enough for realistic energy resolution to not be able to help (this is particularly the case in xenon-based experiments). 
Overall, current experiments have to deal with additional backgrounds to 2$\nu\beta\beta$ and are utilizing additional approaches for their rejection (e.g., pulse shape discrimination, event position, event topology, etc.). Singling out energy resolution underestimates sensitivity. 

Below we briefly summarize the best performing existing experiments, commenting on where they stand in terms of this tree-axis competition.

\subsection{EXO-200} 

EXO-200 (Enriched Xenon Laboratory) experiment is located 
in a salt mine at the Waste Isolation Pilot Plant (WIPP), NM, USA. It has collected low background data from May 2011 until February 2014 (Phase I), when fire and subsequent radioactive waste release interrupted normal activities in the mine. As of February 2016 the experiment has successfully restarted and is planning to continue data taking for another three years (Phase II). The EXO-200 detector is a liquid xenon (LXe) time-projection chamber (TPC). 
EXO-200 has $\sim$200 kg of xenon enriched to $\sim$81\% in $^{136}$Xe (natural abundance $\sim$9\%), with the rest being mostly $^{134}$Xe. 
175 kg are in liquid phase, and 110 kg are in the active volume. The detector is constructed from components carefully selected to minimize radioactive backgrounds~\cite{Leonard:2007uv}. 
The cathode is placed at the center of a cylindrical copper TPC.
Energy depositions in LXe produce both scintillation light ($\sim$176 nm) and charge, which are registered at each end of the TPC by avalanche photodiodes (APDs) and anode wire grids, respectively. 
 Charge deposits in a given event that are spatially separated by $\sim$1 cm or more can be individually resolved. The event can then be classified as single site (SS), or multisite (MS), depending on the number of observed charge deposits. Based on Monte Carlo simulation, $>$90\% of 0$\nu\beta\beta$ events are expected to be reconstructed as SS, while the SS fraction of $\gamma$ events at this energy is $\sim$24\% ($\gamma$ background rejection fraction of roughly 3 to 1). 
Energy of an event is determined by combining the charge and scintillation signals, which achieves better energy resolution than in each channel individually. 
The average energy resolution at the Q-value during Phase-I is $\sim$3.6\% ($\sim$3.9\%) full-width at half-max (FWHM) for SS (MS) events, limited by APD correlated noise. 

Initial 0$\nu\beta\beta$ result was published in 2012~\cite{Auger:2012ar} based on 26.3 kg$\cdot$yr exposure (120.7 live days with fiducial volume containing 79.4 kg $^{136}$Xe). The binned maximum likelihood fit to signal and background components was performed simultaneously for SS and MS events, with background-rich MS dataset constraining residual $\gamma$ contamination of the SS dataset. The analysis was not completely blind, but the data were partially "masked" to hide 2/3 of the live-time for SS events around Q-value to avoid biasing the results. The profile likelihood scan yielded no statistically significant signal, corresponding to the lower limit on the half-life for the light Majorana neutrino mediated decay of T$_{1/2}>$1.6$\cdot$10$^{25}$ yrs at 90\% C.L. The second result~\cite{Albert:2014b} features almost quadrupled exposure (477.6 live days with 76.5 kg $^{136}$Xe fiducial mass, or 100.0 kg$\cdot$yr), addition of standoff distance~\footnote{Distance between a charge deposit and the closest material that is not LXe, other than the cathode.} as observable in the fit, better understood background model, improved reconstruction and energy calibration, better algorithm for scintillation signal extraction that decreased energy resolution, and more detailed systematic error assessment. The best-fit value of 0$\nu\beta\beta$ counts is 9.9, consistent with the null hypothesis at 1.2$\sigma$. The sensitivity of the second analysis is T$_{1/2}>$1.9$\cdot$10$^{25}$ yr, an increase by a factor of 2.7 compared to the first result. The actual limit (T$_{1/2}>$1.1$\cdot$10$^{25}$ yr at 90\% C.L.) is weaker than the first reported, consistent with a statistical fluctuation of background in the ROI around the Q-value (2457.8 keV). 

The best-fit background in $\pm$2$\sigma$ ($\sim$150 keV) ROI is 31.1$\pm$1.8(stat.)$\pm$3.3(sys.) counts, corresponding to the background index (BI) of 250 ROI$^{-1}$tonne$^{-1}$yr$^{-1}$\footnote{Normalized to the exposure with enriched xenon (124 kg$\cdot$yr), not just $^{136}$Xe (100 kg$\cdot$yr).}, consistent with the first result within errors.
The best-fit values of the dominant backgrounds in EXO-200 are external $^{232}$Th (16.0 counts) and $^{238}$U (8.1 counts), and cosmogenically-produced $^{137}$Xe (7.0 counts). The sub-dominant backgrounds are internal $^{222}$Rn, $^{60}$Co summation peak, 2$\nu\beta\beta$ decay, and $\gamma$s from neutron spallation on detector materials. $^{222}$Rn is accurately monitored and tagged using delayed $^{214}$Bi-$^{214}$Po coincidences, suggesting a steady-state population of only $\sim$200 $^{222}$Rn atoms in LXe during normal data taking.
The observed U/Th backgrounds are consistent with estimates from Monte Carlo simulations based on material assay~\cite{exo200_backgrounds}. Higher statistical power of the second analysis suggests a distinct source of $^{232}$Th background from outside of the TPC vessel ("remote" $^{232}$Th). There is also evidence of remote $^{238}$U, but the degeneracy in the shapes of energy and standoff distance distributions do not allow to pinpoint the location, thus not substantiating the suggestion made in the first publication that associated the remote $^{238}$U background with $^{222}$Rn in the air gap between the cryostat and lead shield. The planned operation of a charcoal-based Rn-suppression system during the Phase II data taking may provide additional information. 

For Phase II, EXO-200 is planning several improvements to increase sensitivity. New electronics should decrease APD noise possibly leading to improvement in energy resolution. Flushing the air gap between cryostat and lead shield with Rn-suppressed air should remove or reduce remote $^{238}$U background. Improvements in topology-based background discrimination is expected to increase $\gamma$ background rejection fraction, while maintaining reasonably high signal efficiency. New veto cut for events consistent with neutron capture $\gamma$s on $^{136}$Xe is expected to decrease $^{137}$Xe background. Altogether a factor of 2-3 increase in sensitivity is anticipated after 2-3 years of additional data taking. 

EXO-200 also reported on a search for various Majoron-emitting modes of 0$\nu\beta\beta$ decay~\cite{EXOMajoron:2014}. A lower limit of T$_{1/2}>$1.2$\cdot$10$^{24}$ yr at 90\% C.L. on the half-life of the spectral index = 1 Majoron decay was obtained, corresponding to a constraint on the Majoron-neutrino coupling constant of \(|\langle g^M_{ee}\rangle| < (0.8-1.7)\cdot10^{-5}\).

\subsection{KamLAND-Zen} %

KamLAND-Zen (KamLAND Zero-Neutrino Double-Beta Decay) is located under mt. Ikenoyama near Kamioka, Japan. 
The KamLAND-Zen is a modification of the existing KamLAND detector. 
The KamLAND detector consists of a balloon with 1 kton ultra-pure liquid scintillator (LS). 
The scintillation light is detected by photomultiplier tubes (PMTs) mounted on the stainless-steel containment tank, providing 34\% photocathode coverage. 
The containment tank is filled with non-scintillating mineral oil, shielding the LS from external radiation. The containment tank is surrounded by a 3.2 kton water-Cherenkov detector for cosmic-ray muon identification. KamLAND-Zen consists of 13 tons of Xe-loaded liquid scintillator (Xe-LS) contained in a transparent 3.08 m diameter 25 $\mu m$ thick nylon-based inner balloon (IB), suspended at the center of the KamLAND detector by film straps. 
The KamLAND's main LS acts as a shield for external $\gamma$s and as a detector for internal radiation from the Xe-LS and IB. 
The Xe-LS contained $\sim$2.4 wt.\% of enriched xenon (Phase I), increased to $\sim$2.9\% in 2013 (Phase II). 
The isotopic abundances in the enriched xenon were measured by residual gas analyzer to be $\sim$91\% $^{136}$Xe and $\sim$8.9\% $^{134}$Xe. 
The light yield of the Xe-LS is 3\% lower than that of the main scintillator. The energy resolution at the Q-value is 9.9\% FWHM. 

In spite of the lower resolution and absence of event topology information the experiment currently has the largest amount of isotope of interest ($\sim$320 kg of enr.Xe initially, increased to $\sim$390 kg in 2013) and uniquely large and pure shield of external backgrounds (main KamLAND detector) that provides both passive and active rejection. Initial result was published in 2012~\cite{KZ1} based on 27.4 kg$\cdot$yr exposure (77.6 days with 129 kg of $^{136}$Xe fiducial) and revealed substantial peak in the ROI. The energy of the peak was 3\% larger than the Q-value of $^{136}$Xe $\beta\beta$ decay, excluding the 0$\nu\beta\beta$ decay explanation at more than 5$\sigma$ and thus suggesting a presence of an unexpected background. The background was identified as $^{110m}$Ag, whose presence may be explained by the spallation of $^{136}$Xe by cosmic rays (Xe gas was enriched in Russia and sent to Japan by airplane), or by fallout from the Fukushima I reactor accident on March 11, 2011 (it was observed with Ge detectors in the soil sample in Sendai)~\cite{Gando:2012jr}. The binned maximum likelihood fit was performed on the data, allowing $^{110m}$Ag and other potential unexpected backgrounds to float unconstrained. The resulting 0$\nu\beta\beta$ half-life limit was T$_{1/2}>$5.7$\cdot$10$^{24}$ yr at 90\% C.L. The combined best-fit background rate, dominated by $^{110m}$Ag, was reported in 800 keV ROI (2200-3000 keV) as 0.22 ROI$^{-1}$(Xe-LS tonne)$^{-1}$day$^{-1}$, which we translate using 2.44 wt.\% loading into $\sim$3000 ROI$^{-1}$tonne$^{-1}$yr$^{-1}$ for exposure with enriched xenon.   
The first attempt to remove the background was performed in February 2012 by passing the Xe-LS through a filter. The subsequent data taking showed reduction of $^{110m}$Ag consistent with natural decay of the isotope, indicating that the Xe-LS filtration had not effect. The combined analysis of the datasets before and after the filtration spanned the period from October 2011 to June 2012 (Phase I), corresponding to 89.5 kg$\cdot$yr exposure (112.3 days with 179 kg of $^{136}$Xe before and 101.1 days with 125 kg after the filtration). The 90\% C.L. lower limit of T$_{1/2}>$1.9$\cdot$10$^{25}$ yr was reported~\cite{Gando:2012zm}.

The second attempt to remove $^{110m}$Ag background started in June 2012 when Xe was extracted from the detector and purified. $^{110m}$Ag was confirmed to remain in the depleted LS, which was then also purified. 
The purification eventually reduced $^{110m}$Ag background by more than a factor of 10. Apart from sub-dominant remaining contamination with $^{110m}$Ag, the primary backgrounds for KamLAND-Zen are $^{214}$Bi on the IB film (possibly due to dust contamination during IB film assembly and air leakage during the LS purification), $^{10}$C muon spallation product, and 2$\nu\beta\beta$ decay.
The Phase II dataset was collected between December 2013 and October 2015. The result features increased Xe concentration up to $\sim$2.9 wt.\% for a total Phase-II exposure of 504 kg$\cdot$yr (534.5 days with $\sim$345 kg of $^{136}$Xe), reduction of $^{10}$C background by additional cuts on muon-induced neutron events, and mitigation of the $^{214}$Bi on IB by optimizing fiducial volume cut and addition of volume as an observable in the fit. The energy resolution increased to $\sim$11\% FWHM due to increased number of dead PMTs, compared to Phase I. The Phase II dataset is further divided into two equal periods (Period-1 and Period-2) roughly equal to one lifetime of $^{110m}$Ag each. The fit is simultaneous in energy and volume (with 20 equal volume bins) and is performed independently for the two time periods.
The best-fit background index in 400 keV ROI (2300-2700 keV) for the cleanest part of the dataset (Phase II, Period-2, 1-m radius spherical volume) translates to $\sim$160 ROI$^{-1}$tonne$^{-1}$yr$^{-1}$ for exposure with enriched xenon. 
The Phase II result is T$_{1/2}>$9.6$\cdot$10$^{25}$ yr at 90\% C.L. Combining Phase I and Phase II data gives a 90\% C.L. lower limit of T$_{1/2}>$1.1$\cdot$10$^{26}$ yr at 90\% C.L.~\cite{KZ_new}. This result solidifies the KamLAND-Zen's lead among current generation experiments.
KamLAND-Zen's plan for Phase III is to rebuild the IB with cleaner material, increase xenon amount to $\sim$800 kg, refurbish water-Cerenkov veto detector. The projected sensitivity for Phase III is T$_{1/2}>$2$\cdot$10$^{26}$ yr after two years of data taking\cite{KZ_Phase2}.

KamLAND-Zen also reported on searches for various Majoron-emitting modes of 0$\nu\beta\beta$ decay using 38.6 kg$\cdot$yr~\cite{KamLANDZen:2012zzg} and on 0$\nu\beta\beta$ decays to the excited states using 89.5 kg$\cdot$yr~\cite{KZ_excited} exposures. A lower limit on the ordinary (spectral index n = 1) Majoron-emitting decay half-life of $^{136}$Xe was obtained as T$_{1/2}>$2.6$\cdot$10$^{24}$ yr at 90\% C.L. The corresponding limit on the Majoron-neutrino coupling constant was reported as \(|\langle g^M_{ee}\rangle| < (0.8-1.6)\cdot10^{-5}\). The authors of this manuscript, however, believe that the coupling constraint should be a factor of two stronger, due to the improved understanding of the phase space factor value for the n=1 Majoron decay~\cite{EXOMajoron:2014}. The established lower half-life limits for the 0$^+_1$, 2$^+_1$, and 2$^+_2$ state transitions are T$_{1/2}>$2.4$\cdot$10$^{25}$, $>$2.6$\cdot$10$^{25}$ yr, and $>$2.6$\cdot$10$^{25}$ yr at 90\% C.L., respectively.

\subsection{Gerda}

The Germanium Detector Array (GERDA) uses high-purity germanium detectors enriched in the isotope $^{76}$Ge. $^{76}$Ge has a Q-value = (2039.061 $\pm$ 0.007) keV\cite{Ge_qvalue} and a natural abundance of 7.73\%. The advantage of this technology is that it provides the best energy resolution of any detector practical for use in rare events searches. However, one of the major drawbacks is the difficulty of the enrichment, as germanium is a solid at room temperature (whereas xenon is a gas and tellurium experiments don't need to enrich). Growing the crystals and the complicated process of purifying germanium combine to make the technique difficult to scale in mass. With that said, germanium based experiments have found several techniques to achieve very-low backgrounds and remain competitive.

Phase I of GERDA uses reprocessed p-type coaxial detectors that were formerly owned by Heidelberg-Moscow (HDM) and the International Germanium Experiment (IGEX). The total mass of the eight reprocessed detectors (5 from HDM, 3 from IGEX) in Phase I is 17.67 kg enriched to $\sim$86\% in $^{76}$Ge. These detectors achieve an interpolated FWHM energy resolution of between 0.20\% and 0.28\% (4.2 and 5.7 keV) at Q-value. The detectors operate at the Italian INFN Laboratori Nationali del Gran Sasso (LNGS). The experiment is shielded by 3 m of water instrumented with PMTs to act as a muon-veto and a liquid argon (LAr) shield. The LAr shield also offers cooling but comes with a drawback that $^{42}$Ar which decays within the LAr becomes $^{42}$K. These $^{42}$K daughters are often charged and thus drift toward the detectors, leading to a background from the $^{42}$K decay on the detector surface. However, in Phase II of the experiment this background will be greatly reduced. The detectors are held by low-mass copper supports in strings. They have demonstrated in an \textit{ex situ} test that that $^{42}$K background can be eliminated by an external shroud around the detector. In Phase-II, they will use a shroud made from low-background nylon. A pulse-shape discrimination (PSD) based on an artificial neural network rejects multiple scatter events (45\% of the background) with a signal acceptance of 0.90$^{+0.05}_{-0.09}$.

For Phase II, GERDA is planning to use broad energy germanium (BEGe) detectors manufactured by Canberra\cite{gerda_detector}. BEGe detectors are not coaxial and instead use a point contact to reduce capacitance and gain greater multiple scatter rejection. The BEGe detectors show an improved FWHM energy resolution of between 0.13\% and 0.2\% at Q-value
 (30-40 rel.\% better than p-type coaxial detectors) and simpler PSD using an amplitude (A) over energy (E) cut. The A/E cut in BEGe detectors rejects 80\% of the background while keeping a fraction 0.92 $\pm$ 0.02 of the signal.

GERDA's first result is based on the eight detectors from HDM and IGEX deployed in November 2011 and three BEGe detectors deployed in July 2012\cite{Gerda:2013}. Two coaxial detectors, which started to draw leakage current, and one BGE detector, which showed an unstable behavior, are omitted in the analysis. All events in $\pm$20 keV window around Q-value were blinded until all cuts and analysis were finalized. A total of 21.6 kg$\cdot$yr of data are collected with three events in $\pm$5 keV of Q-value passing all cuts, placing a 90\% C.L. limit of T$_{1/2}>$2.1$\cdot$10$^{25}$ yr at 90\% C.L. Combining this result with previous results in germanium yields a limit of T$_{1/2}>$3.0$\cdot$10$^{25}$ yr at 90\% C.L., inconsistent with the claimed discovery by part of HDM. 
The observed background was found to be consistent within errors with prediction~\cite{gerda_backgrounds}, which has flat energy spectrum around Q-value and BI for 5 keV ($\sim$FWHM) wide ROI, normalized to exposure with enr.Ge, of $\sim$50 ROI$^{-1}$tonne$^{-1}$yr$^{-1}$ after PSD cuts (based on rate per keV quoted in~\cite{Gerda:2013}). Dominant components are $^{42}$K, $^{214}$Bi, $^{228}$Th, $^{60}$Co, and $\alpha$ emitting isotopes in $^{226}$Ra chain. 
In the future GERDA intends to integrate a total of 20 kg of BEGe detectors which will be closely packed to increase vetoing power of multiple scatter events. Furthermore, the LAr shield is being instrumented as an active veto. In one test facility a rejection of 99.9\% of all $^{42}$K backgrounds events has been demonstrated. The overall goal is a total background reduction of one order of magnitude over the Phase I detectors and eventual sensitivity of T$_{1/2}>$10$^{26}$ yr at 90\% C.L.

GERDA also reported on a search for various Majoron-emitting modes of 0$\nu\beta\beta$ decay~\cite{gerda_majoron}. A lower limit of T$_{1/2}>$4.2$\cdot$10$^{23}$ yr at 90\% C.L. on the half-life of the spectral index = 1 Majoron decay was obtained, corresponding to a constraint on the Majoron-neutrino coupling constant of \(|\langle g^M_{ee}\rangle| < (3.4-8.7)\cdot10^{-5}\).

\subsection{CUORE-0}

The Cryogenic Underground Observatory for Rare Events (COURE) takes place at LNGS. COURE will eventually consist of 19 towers containing 52 $^{nat}$TeO$_{2}$ bolometers each. The first stage is a single tower referred to as COURE-0, which collected data from March 2013 to August 2013 and November 2013 until March 2015. Each crystal is cooled down to $\sim$10 mK and uses a neutron-transmutation-doped Ge thermistor to measure small changes in temperature. When an ionizing particle interacts in a given crystal, that creates a pulse of heat which is detected by the thermistor. This technique has several distinct advantages. 
First, the isotope of interest, $^{130}$Te, has a natural abundance of $\sim$34\% meaning that isotopic enrichment is unnecessary. 
Second, the energy resolution achievable with this technology is similar to germanium detectors, $\sim$5.1 keV FWHM at Q-value. However the need to run at 10's of mK adds cost to the experiment per unit size. Also the events last for on the order of seconds, so event pile-up can be a concern. But most seriously, because there is a limit to how large the crystals can be, there are risks of backgrounds from surface contamination. Part of the purpose of COURE-0 was to demonstrate the ability to sufficiently reduce these backgrounds from surface events.

A rigorous program for growth of radiopure crystals and surface cleaning has been developed to address backgrounds~\cite{cuore_crystals}. The cryogenic systems, shielding, and electronics are all reused from the predecessor experiment, Couricino~\cite{cuorcino_1,cuorcino_2,cuorcino_3}. The experiment is shielded by Roman lead inside the cryostat. As compared to Couricino, the amount of copper per unit TeO$_{2}$ has been largely reduced to mitigate $^{232}$Th backgrounds. Overall, BI has been substantially reduced compared to Couricino (see Table 4 in~\cite{cuorcino_3}).

In the first analysis, COURE-0 implemented a data-salting scheme to effectively blind the ROI. A random fraction of 1-3\% of the events within $\pm$10 keV of Q-value and events within $\pm$10 keV of the 2.615 MeV line of $^{208}$Tl are switched. The fractions varies run-to-run, and creates a peak at Q-value. The analysis exploits both optimal filtering and decouple optimal filtering to separate signal from noise in measuring pulse amplitude. There are also two methods of gain stabilization, based either on the heater data or calibration data. For each crystal-dataset, the combination of filtering and gain stabilization that optimizes energy resolution is utilized. This does vary dataset-to-dataset. Backgrounds are further reduced in the analysis based on six pulse shape parameters. 50\% of the data is randomly selected, excluding the ROI, and prominent peaks from the background are used to tune cuts to maximize the ratio of the signal to the square root of the background. 
The overall efficiency is (81.3 $\pm$ 0.6)\%. 
There are some uncertainties in the calibrated energy scale observed for different peaks. Using the sum peak at 2505.7 keV from $^{60}$Co, a shift in the measured energy of 1.9 keV was observed (so that the reconstructed energy of the peak was 2507.6 keV). This is due to a difference in detector response between two-gamma events and single gamma events which is still under investigation. As a result, an uncertainty in the reconstructed Q-value is taken based on residuals at prominent peaks in the background spectrum.

After unblinding, a total of 233 candidate events passed all cuts in the 100 keV region (2.47-2.57 MeV) around Q-value with an exposure of 35.2 kg$\cdot$yr of natural Te and 9.8 kg$\cdot$yr of $^{130}$Te. After applying a profile likelihood analysis and accounting for all systematics, a 90\% C.L. limit of 2.7$\cdot$10$^{24}$ yr is placed on T$_{1/2}$~\cite{cuore0}. Combing this with the previous result from Couricino gives a 90\% C.L. limit of  T$_{1/2}>$4.0$\cdot$10$^{24}$ yr. The reported best-fit BI translates to $\sim$300 ROI$^{-1}$tonne$^{-1}$yr$^{-1}$ for 5.1 keV ROI, normalized to the exposure with natural Te.

The next stage is to implement the remaining 18 towers of CUORE, for a total mass of 741 kg TeO$_{2}$ or 206 kg of $^{130}$Te. This will allow for further background reduction by vetoing events coincident between multiple detectors. CUORE will also have additional external shielding. CUORE has a projected sensitivity to T$_{1/2}>$ 9.0$\cdot$10$^{25}$ years.

\subsection{Others}

Several other efforts do not yet have comparable results but are working on promising technologies. 

NEXT ("Neutrino Experiment with a Xenon TPC") is planning to use high-pressure xenon gas (gXe) TPC with electroluminescent (EL) readout to achieve background rejection substantially better than other xenon detectors. At 15 bar pressure, a $\beta\beta$ decay typically leaves a $\sim$15 cm length track in gXe (compared to $\sim$2 mm in LXe). A characteristic increase of ionization density at both ends of the track - two "blobs" - can be used to effectively reject the two major backgrounds, $^{208}$Tl ($^{232}$Th chain) and $^{214}$Bi ($^{238}$U chain), whose tracks should typically have just one blob due to a single electron. This should in principle allow to identify background (signal) events virtually indistinguishable from signal (background) in a conventional LXe detector - the $\sim$3\% of $\gamma$ depositions at energies near the Q-value ($^{214}$Bi) occurring via the photoeffect (no Compton scatters) and thus having true SS topology, and several percent of $\beta\beta$ events that emit Bremsstrahlung photons and thus having true MS topology. 
Moreover, gaseous xenon offers an order of magnitude better energy resolution than LXe due to small Fano factor, hence further suppressing backgrounds, including the otherwise irreducible 2$\nu\beta\beta$ decay. A  challenge of this approach is reaching large exposure due to low density of gas and practical limits of building and operating high-pressure detectors. NEXT collaboration makes steady progress through a series of smaller scale prototypes. The first two $\sim$1 kg scale prototypes, called NEXT-DBDM and NEXT-DEMO, have demonstrated an extrapolated FWHM energy resolution of 0.5\% at Q-value~\cite{next_dbdm} and an existence of topological signature~\cite{next_demo}, respectively. The collaboration is constructing a $\sim$50 kg scale prototype, NEW (NEXT-WHITE), at the Canfranc underground laboratory, Spain, to validate the design and background model for the upcoming factor of two upgrade (NEXT-100). The installation and commissioning of NEXT-100 is planned for 2017. Based on Monte-Carlo prediction of BI of $<$4$\cdot$10$^{-4}$ keV$^{-1}$kg$^{-1}$yr$^{-1}$, the projected sensitivity of NEXT-100 is T$_{1/2}>$6$\cdot$10$^{25}$ yr at 90\% C.L. after three years of running~\cite{next_sens}, still giving it a chance to catch-up with the field. 

A similar approach is pursued by PandaX-III collaboration, which also plans to use high-pressure gXe. The distinct feature is not using light information for either energy or tracking. Instead a micropatterned charge readout (using micromega detectors) is expected to provide good energy and spatial resolution with even lower background~\cite{pandax}. The goal is to first build a 200 kg gXE TPC, with the installation in the China Jinping Underground lab expected by 2017. The successful demonstration of this modular approach may provide a straightforward and attractive path towards the tonne-scale experiment. 

NEMO ("Neutrino Ettore Majorana Observatory") is another program that offers superior tracking capabilities, but faces greater challenge with exposure. The previous generation experiment, NEMO-3, searched for 0$\nu\beta\beta$ decay of seven isotopes from February 2003 to January 2011. The NEMO-3 detector is located in the Modane underground laboratory, France, and is capable of reconstructing full topology of events (tracks of individual $\beta$s and their energy). This not only offers greater background rejection, but may also allow to discern different mechanisms of 0$\nu\beta\beta$ decay. The NEMO-3 detector consists of sectors arranged in a cylindrical geometry containing thin source foils of $\beta\beta$ emitters. The foils are suspended between two concentric cylindrical tracking volumes consisting of drift cells operating in Geiger mode~\cite{nemo3_nim}. The tracking detector, immersed in a magnetic field, is surrounded by a plastic scintillator calorimeter, monitored by PMTs. Combination of tracking and calorimetry allows to distinguish $\beta$ and $\gamma$ events, while magnetic field allows one to reject pair production and external electron events. The detector is shielded from external $\gamma$ backgrounds by iron and water with boric acid to suppress the neutron flux. A radon trapping facility reduces internal radon background. The strongest result was obtained for $^{100}$Mo, for which 34.7 kg$\cdot$yr exposure was accumulated. No evidence for the 0$\nu\beta\beta$ signal has been found, yielding a limit for the light Majorana neutrino mass mechanism of T$_{1/2}>$1.1$\cdot$10$^{24}$ years at 90\% C.L.~\cite{nemo3-mo}. In spite of great advantages of this technology, the challenges are smallness of target foils (only 6.9 kg of $^{100}$Mo used by NEMO-3), not great energy resolution ($\sim$9\% FWHM at 3034.4 keV, the Q-value of $^{100}$Mo) and small detection efficiency (11.3\% for 0$\nu\beta\beta$ decay of $^{100}$Mo for $\beta\beta$ energy above 2 MeV). The current stage of the program, Super-NEMO, aims to reach sensitivity up to T$_{1/2}>$10$^{26}$ yr~\cite{supernemo}, rivaling that of KamLAND-Zen's Phase III. It plans to achieve that by increasing mass to 100 kg, using $^{82}$Se as primary isoptope (due to 13 times smaller rate of 2$\nu\beta\beta$ background that with $^{100}$Mo), improving energy resolution to 4\% FWHM at 3 MeV (using bigger PMTs with higher quantum efficiency and no light-guides), and decreasing Th/U contamination (by material selection, target isotope purification, and radon trapping). Commissioning of the first module, the SuperNEMO demonstrator, is expected at the end of 2016.

Majorana Demonstrator is an $^{76}$Ge experiment under construction in the Sanford Underground Research Facility (SURF). It is complementary to GERDA and plans to join in efforts for the next generation 0$\nu\beta\beta$ search. Its main goal is to demonstrate a path towards sensitivity covering IH, which requires a BI in the 4 keV ROI around Q-value of $\leq$3 ROI$^{-1}$tonne$^{-1}$yr$^{-1}$, which should scale to 1 ROI$^{-1}$tonne$^{-1}$yr$^{-1}$ in a tonne-scale experiment due to better self-shielding. The signature feature of Majorana is the ultra-low background copper used in construction of the cryostat and other detector components. The copper is made by the collaboration using electroforming at SURF and shallow underground site of PNNL. Using the most sensitive ICP-MS-based assay technique~\cite{majorana_icpms}, the copper was shown to have U/Th limits of $<$0.1 $\mu$Bq/kg. In addition to the GERDA's PSD analysis, Majorana also developed an alternative approach~\cite{majorana_psd} for discriminating multi- and single-site events. Analysis of natural background data (in particular, single and double escape events of $^{208}$Tl, which have MS and SS topology, respectively) resulted in a survival probability of just 1.0($\pm$0.2)\% for such MS events versus 98.3($\pm$21.9)\% of SS events, roughly corresponding to 100 to 1 rejection fraction. The first full detector module containing 20 enriched (16.8 kg) and 9 natural (5.7 kg) Ge detectors was installed and started in-shield measurements in June 2015. A second module with another 15 (12.9 kg) enriched and 15 (9.4 kg) natural Ge detectors was being assembled at the end of 2015. If the expected BI of 3 ROI$^{-1}$tonne$^{-1}$yr$^{-1}$ (about an order of magnitude smaller than GERDA's Phase I) is achieved, Majorana Demonstrator will reach sensitivity level of 1$\cdot$10$^{26}$ yr after about three years of operation~\cite{majorana_demonstrator}.

The AMoRE (Advanced Mo based Rare process Experiment) collaboration is going to use $^{40}$Ca$^{100}$MoO$_4$ (CMO) crystals with a total mass of 100 kg as cryogenic scintillation detector to search for 0$\nu\beta\beta$ decay of $^{100}$Mo~\cite{amore}. Simultaneous detection of phonons and scintillation light is a potentially very powerful tool to reject backgrounds from internal radioactive contamination of crystals. The collaboration is making progress in development of large crystals and characterization of their radio-purity and optical properties~\cite{amore}. The energy resolution was demonstrated to be comparable to that of germanium detectors~\cite{amore2}. The first phase (AMoRE-I) will use 5 kg (possible, 10 kg) of CMO crystals to be installed at the YangYang underground laboratory, South Korea. The aimed background level in $\pm$10 keV ROI around $^{100}$Mo Q-value (3.034 MeV) is 40 ROI$^{-1}$tonne$^{-1}$yr$^{-1}$ (based on rate per keV quoted in~\cite{amore3}).  AMoRe is expected to eventually achieve sensitivity of 3$\cdot$10$^{26}$ yr after 250 kg$\cdot$yr exposure.

\subsection{Comparison of current results and combined limit}

\begin{figure*}[htp]
\centering
\includegraphics[scale=0.25]{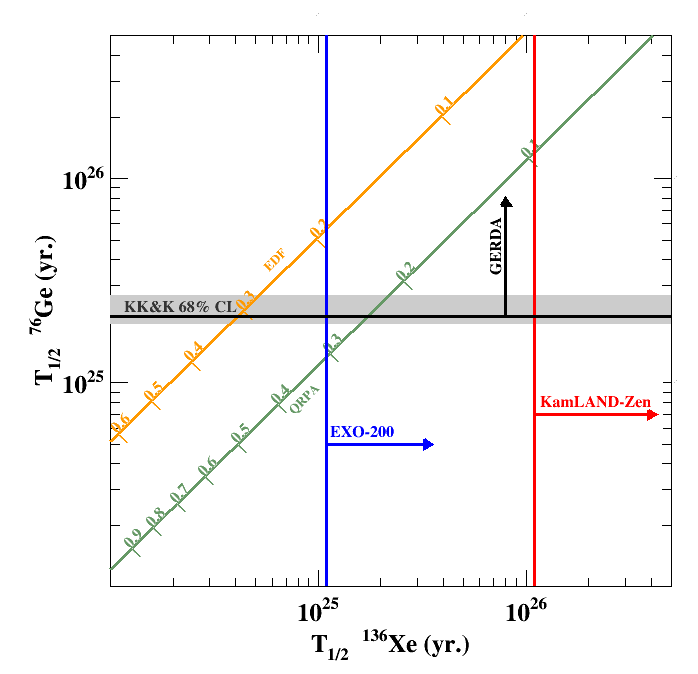}
\includegraphics[scale=0.25]{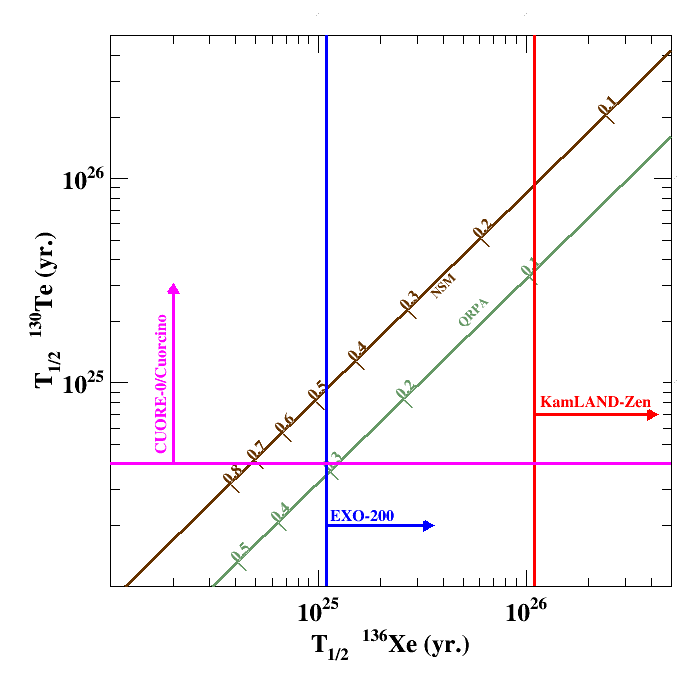}
\includegraphics[scale=0.25]{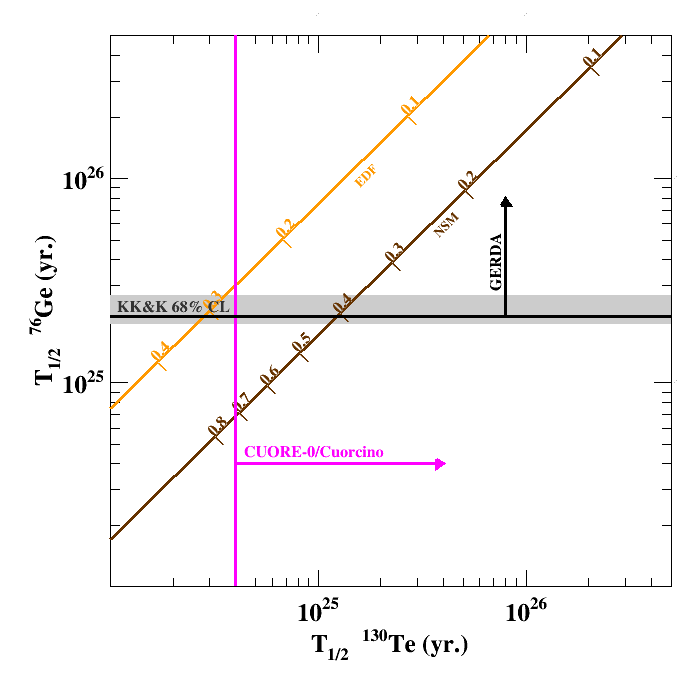}
\vspace*{8pt}
\caption{\label{fig:GrattaVogel} Comparisons of 0$\nu\beta\beta$ 90\% C.L. half-life limits and implications for effective Majorana neutrino mass for Ge versus Xe, Xe versus Te, and Ge versus Te. The diagonal axes represent different nuclear matrix element models (EDF~\cite{edf}, NSM~\cite{nsm}, QRPA~\cite{qrpa}), with the tick marks showing the relevant mass limit in eV. Unquenched value of QRPA from~\cite{qrpa} is used. NSM value averaged over the two short range correlation parameterizations used in~\cite{nsm} is used. Phase-space factors taken from~\cite{iachello_psf}.} 
\end{figure*}

Table~\ref{tab:results} summarizes half-life limits for light Majorana mediated 0$\nu\beta\beta$ decay and BIs for the currently leading experiments. The BIs are expressed in units of ROI$^{-1}$(isotope tonne)$^{-1}$yr$^{-1}$. Note the unusual BI normalization in the table - per mass of the isotope of interest, instead of per mass of the element (as is typically done by experiments and also by us in the detailed descriptions above). We think that normalizing per mass of the isotope of interest provides for the more straightforward comparison and insight into how far the present results are from the next generation goal of covering IH. In order to compare the half-life results obtained with different isotopes we need to convert them into limits on effective Majorana mass. Fig.~\ref{fig:GrattaVogel} shows such comparisons assuming the simplest mechanism of 0$\nu\beta\beta$ decay and vacuum value of the axial vector coupling constant. The range of limits on the effective Majorana mass depends on used NMEs (EDF~\cite{edf}, NSM~\cite{nsm}, (R)QRPA~\cite{qrpa}) and phase-space factors~\cite{iachello_psf}.  
\begin{savenotes}
\begin{table}[ht]
\tbl{Summary of comparison of leading 0$\nu\beta\beta$ decay experiments. BIs are normalized per mass of the isotope of interest.}
{\begin{tabular}{@{}p{21.mm}p{10mm}p{20mm}p{20mm}p{19mm}p{17.5mm}@{}} \toprule
Experiment & Isotope mass, tonne & T$_{1/2}$ 90\% C.L., 10$^{25}$ yr & m$_{\beta\beta}$ 90\% C.L., eV & BI, ROI$^{-1}$(isotope tonne)$^{-1}$yr$^{-1}$ & ROI, keV \\
\colrule
EXO-200 & 0.16 & 1.1 & 0.19-0.47 & 310 & 150 ($\pm$2$\sigma$) \\
KamLAND-Zen & 0.32 & 11 & 0.06-0.15 & 180 & 400 \\
GERDA & 0.018 & 2.1 & 0.24-0.41 & 58 & 5 (FWHM) \\
CUORE-0 & 0.013 & 0.4 & 0.26-0.71 & 890 & 5.1 (FWHM)\\ \botrule
\end{tabular}\label{tab:results} }
\end{table}
\end{savenotes}

The combination of results from several datasets (Cuorcino and Cuore-0, NEMO-3, first phases of EXO-200, KamLAND-Zen, and GERDA) was recently reported~\cite{combination} as $m_{\beta\beta}<$130-310 meV at 90\% C.L. 
The combined result, however, would be completely dominated by KamLAND-Zen if its latest result~\cite{KZ_new} is taken into account.

\section{Towards inverted hierarchy}

While the results of the currently running experiments will likely continue to be iteratively updated in the next couple of years, the competition between them is, fundamentally, completed. The existing favorites that have already produced strong results in qausi-degenerate are using the momentum to lead the field towards the next logical step - attempting to cover the IH. In spite of the impressive performance, experiments need to further reduce backgrounds and provide the necessary increase in exposure. Some approaches that are taking longer to get data promise better radiopurity or background rejection and may still surmount currently leading technologies. Below we describe currently active tonne-scale efforts with the established goal of exploring the IH. 
 
\subsection{nEXO} 

nEXO ("next EXO") is being designed as a 5 tonne enriched LXe cylindrical TPC detector. It borrows basic approach from EXO-200, but several improvements need to be made to achieve the goal of having 3$\sigma$ discovery potential completely covering the IH region after 10 years of data. Having a bigger homogeneous detector allows one to take a better advantage of external $\gamma$ background reduction by self-shielding. To remove any material from the most pure central region of the detector, nEXO is planning to place the cathode at an end of the TPC, instead of in the middle, as EXO-200. The challenge here is to maintain low loss of ionization electrons to impurities during the longer drift length, which is expected to be achievable with increased xenon re-circulation flow rate, and to demonstrate stable operation with larger high voltage needed to maintain the electric field in the bigger detector. nEXO needs to improve the energy resolution compared to what has been demonstrated with EXO-200 to keep 2$\nu\beta\beta$ decay from becoming a significant background. It expects to achieve this by increasing the light collection and reducing noise, by covering up to $\sim$4-5 m$^2$ of the TPC's barrel with silicon photomultipliers~\cite{nEXO_sipm}, which have much higher gain than APDs used in EXO-200, and by maximizing reflectivity of internal components. nEXO is planning to deploy the detector at a deeper location (e.g., SNOLAB) than EXO-200, thus reducing cosmogenically-produced $^{137}$Xe background, which currently amounts to roughly a quarter of the background budget in EXO-200. In EXO-200 resolving multiple scatters allowed for roughly 3 to 1 rejection of $\gamma$ backgrounds at Q-value. nEXO aims to double the rejection fraction with the help of new charge readout scheme, now under prototyping, and lower noise electronics, possibly submerged in LXe. A continuing challenge is the radiopurity of detector components. For example, nEXO needs to demonstrate an order of magnitude lower limits on U/Th backgrounds in TPC copper, compared to EXO-200, in order for the projected discovery potential to completely cover the IH region (for the worst-case NME and no g$_A$ quenching). The collaboration is expanding the screening and characterization program to address the radiopurity questions. An ultimate background rejection technique would be to identify individual $^{136}$Ba atoms or ions following $\beta\beta$ decay of $^{136}$Xe ("Ba-tagging"). While work on various approaches to Ba-tagging continues~\cite{ba_karl,ba_brian,ba_thomas}, at least initial operation of nEXO is not expected to achieve it. Xenon is one of the best choices of isotopes in terms of scalability, but amassing 5 tonne of enriched isotope is not trivial. World production of xenon in 1998 was estimated to be 28-40 tonne/yr~\cite{xenon_production}, which corresponds to 3-4 tonne of $^{136}$Xe.

\subsection{LZ} 

The LUX-ZEPLIN (LZ) experiment is an ultra-low-background two-phase natural xenon TPC whose main goal is to search for WIMP dark matter with 7 tons of active mass. However, as a result LZ will have some sensitivity to 0$\nu\beta\beta$. Major background contributions include the TPC PMTs, the xenon vessel, and the resistors in the field cage and cathode high voltage connection. There is some contribution from the radioactivity of the Davis cavern walls that is still under investigation. At worst some additional shielding, over what is in the current baseline plan, may be necessary above and below the xenon vessel, but not on the sides. In order to extend the dynamic range over previous experiments, LZ will have two DAQ channels, one with high-gain for the low-energy (0 to 100's of keV) data and one with low-gain for high-energy ($\sim$ MeV) data. The LZ projection is a median expected 90\% C.L. sensitivity of T$_{1/2}>$ 1.0$\cdot$ 10$^{26}$ yr. LZ is planned to start taking data in 2020 and will run for five years. In the future, this half-life sensitivity could be extended by one order of magnitude by either filling LZ with enriched xenon or in a similar 50 tonne detector filled with natural xenon. LZ is already fully funded, so doing the former can be accomplished at only a fraction of the cost of a dedicated tonne-scale LXe experiment, providing for a natural synergy between the dark matter and 0$\nu\beta\beta$ research goals. In a 50 tonne detector, one would exploit the self-shielding of liquid xenon to reduce backgrounds. The DARWIN collaboration has made a similar suggestion~\cite{darwin}.

\subsection{KamLAND2-Zen} 

As discussed in the previous section, KamLAND-Zen is likely to continue being the most sensitive experiment in the immediate future. To compete with the next generation experiments, however, several challenges need to be addressed by the future detector upgrade, called "KamLAND2-Zen". The amount of enriched xenon will be further increased (in excess of 1 tonne), with a target sensitivity fully covering IH in a 5 yr measurement. Energy resolution needs to be improved, as this is the only discriminator against 2$\nu\beta\beta$ background. The three-pronged approach is expected to improve FWHM resolution at Q-value from $\sim$9.9\% to $<$6\% by using light concentrators, brighter scintillator, and PMTs with higher quantum efficiency. Additional background rejection is foreseen with an imaging system and a scintillating balloon film that are being developed~\cite{KZ_Phase2}. A technique to separate scintillation and Cerenkov light to reconstruct direction
of electrons using ultra-fast light detectors is being investigated~\cite{kz_lindley} and may lead to further improvements in background rejection. 

\subsection{SNO+} 

SNO+ is a large liquid scintillator experiment that will use $^{130}$Te (Q-value=2527.5 keV). Tellurium offers very good scalabity due to high natural abundance (34\%) of $^{130}$Te, not requiring enrichment. SNO+ will fill 6-m radius spherical acrylic vessel (AV) located in the SNOLAB, Canada, with about 0.78 kton of liquid scintillator. The scintillator will be loaded with Te. Initial concentration is set to 0.3\% (800 kg of $^{130}$Te), which was demonstrated to retain good light yield, attenuation, and 2 year stability~\cite{sno}. The scintillator's purity is expected to be comparable to KamLAND's. The scintillation time profile depends on the density of energy depositions, allowing one to discriminate $\alpha$ backgrounds. The scintillation light will be viewed by $\sim$9300 PMTs supported by a geodesic stainless steel structure, filled with about 7 kton of ultra-pure water, which provides a shield for external backgrounds. Further external background rejection is provided by self-shielding. Internal backgrounds could be measured \textit{in situ} prior to $^{130}$Te loading to address possible presence of unexpected sources. Expected main internal backgrounds, in the order of contribution, are $^8$B solar neutrinos (flat continuum from elastically scattered electrons), 2$\nu\beta\beta$,  and $^{238}$U/$^{232}$Th chains. Sub-dominant sources are cosmogenically produced isotopes and ($\alpha$,n) reaction with subsequent release of 2.22 MeV neutron capture $\gamma$. Assuming 3.5 m fiducial volume cut and $\sim$10\% FWHM energy resolution at Q-value, the BI of $\sim$27 ROI$^{-1}$tonne$^{-1}$yr$^{-1}$ is anticipated (with ROI defined from $-$0.5$\sigma$ to $+$1.5$\sigma$ around the Gaussian signal peak). SNO+ has projected sensitivity of 9$\cdot$10$^{25}$ yr for 0.3\% loading after 5 years (Phase I). Phase I is expected to start in 2017. The challenge for covering IH is to increase the isotope loading while maintaining good enough attenuation length, light emission levels, and scintillator stability. Upgrade to high quantum efficiency PMTs and improvements to PMT concentrators should increase the light yield by a factor of $\sim$∼3 while the possibility of an order of magnitude larger isotope loading is under investigation. 

\subsection{Gerda/Majorana tonne scale} 

Gerda and Majorana collaboration have formally agreed to share their expertise and combine the efforts to create the next generation tonne-scale $^{76}$Ge experiment, which will likely be implemented in stages of increasing mass (e.g., 250, 500, and 1000 kg). The experiment will "cherry-pick" and combine best solutions being developed by current generation. In spite of the inherently strong background rejection capability, thanks to great energy resolution and PSD algorithms, demonstrating small enough background level to cover IH is still an outstanding challenge. Amassing a tonne-scale of enriched germanium detectors, ideally produced underground, is also non-trivial. 

\subsection{CUPID/Lucifer} 

As mentioned previously, perhaps the largest limitation to bolometers for 0$\nu\beta\beta$ decay searches is the background from degraded $\alpha$s from surface contamination. As a perfectly clean crystal surface is practically impossible, bolometers which can discriminate $\alpha$ background become extremely attractive. CUORE Upgrade with Particle Identification (CUPID) is a project to develop scintillating bolometers and/or pulse shape discrimination in order to be able to reject $\alpha$ backgrounds~\cite{cupid}. In addition, the CUPID collaboration believes improvements are possible by reducing energy resolution, finding lower radioactivity materials, and isotopic enrichment. The first stage, Low-background Underground Cryogenics Installation For Elusive Rates (LUCIFER), is using 17 kg ZnSe and 14 kg ZnMoO$_{4}$ and plans to start data-taking in 2016. The goal of LUCIFER is a 90\% C.L. median expected limit of T$_{1/2}>$1.8$\cdot$10$^{25}$ yr for $^{82}$Se and T$_{1/2}>$6.2$\cdot$10$^{24}$ yr for $^{100}$Mo after two years. The eventual goal of the CUPID program is sensitivity to the Majorana neutrino mass scale of $\sim$10 meV.

\section{Concluding remarks}

Search for 0$\nu\beta\beta$ decay continues to be one of the most promising directions in neutrino physics. In the US, the 2015 Long Range Plan for Nuclear Science recommends a timely development of a US-led ton-scale 0$\nu\beta\beta$ experiment, suggesting vigorous detector and accelerator R\&D in its support~\cite{nsac}. While the general strategy is unambiguous, some recent developments in nuclear theory may justify the fine-tuning of the near term effort. If the axial vector coupling constant is indeed quenched in 0$\nu\beta\beta$ decay, then the investment in the next generation experiments, foreseen for the next decade, will effectively provide an order of magnitude smaller discovery potential than originally expected. In that case it could be prudent to choose synergistic experiments, which can provide competitive 0$\nu\beta\beta$ reach at a fraction of the cost, while pursuing R\&D into novel technologies that would provide definitive reach under all circumstances. 

\section*{Acknowledgments}
We thank Liang Yang for useful feedback on the manuscript and Petr Vogel for suggesting most up-to-date NMEs and phase-space factors to be used in Fig.~\ref{fig:GrattaVogel}

\bibliographystyle{ws-mpla}
\bibliography{References}

\end{document}